# A Brief Review of Current Lithium Ion Battery Technology and Potential Solid State Battery Technologies

Andrew Ulvestad

**Abstract**

Solid state battery technology has recently garnered considerable interest from companies including Toyota, BMW, Dyson, and others. The primary driver behind the commercialization of solid state batteries (SSBs) is to enable the use of lithium metal as the anode, as opposed to the currently used carbon anode, which would result in ~20% energy density improvement. However, no reported solid state battery to date meets all of the performance metrics of state of the art liquid electrolyte lithium ion batteries (LIBs) and indeed several solid state electrolyte (SSE) technologies may never reach parity with current LIBs. We begin with a review of state of the art LIBs, including their current performance characteristics, commercial trends in cost, and future possibilities. We then discuss current SSB research by focusing on three classes of solid state electrolytes: Sulfides, Polymers, and Oxides. We discuss recent and ongoing commercialization attempts in the SSB field. Finally, we conclude with our perspective and timeline for the future of commercial batteries.

**Introduction**

The demand for better batteries is driven by many industries. Rechargeable lithium ion batteries have emerged as the dominant energy storage source for consumer electronics, automotive, and stationary storage applications. In particular, the $LiCoO_2$ (LCO) cathode and its transition metal oxide successors (notably $LiNi_{0.8}Co_{0.15}Al_{0.05}O_2$ - NCA and $LiNi_xMn_yCo_zO_2$ – NMCxyz with x+y+z=1) coupled with the carbon (C) anode and a liquid electrolyte with additives has been incredibly successful since the early 1990s. However, this chemistry is not without limitations. Pure electric vehicles have yet to achieve cost parity with gasoline cars due in large part to battery cost (estimated as ~37% of the Tesla Model 3 cost, see Appendix) and range anxiety (only 4 pure electric cars have ranges > 200 miles on a single charge). Both of these issues are significantly dependent on battery energy density. Additionally, the liquid electrolyte is flammable, hazardous, and is a significant cost/weight in the battery. Naturally, these problems continue to motivate the search for cheaper, higher energy density battery chemistries.

One obvious change to the oxide cathode/C anode paradigm is to replace the C anode with Li metal ($Li_m$). The $Li_m$ anode has approximately 10x the gravimetric capacity (Ah/g) of the C anode. While this does not translate into a 10x improvement in gravimetric energy density, as we discuss later, the gains are potentially significant. However, repeated charge/discharge cycles leads to non-uniform stripping and deposition of $Li_m$, leading to dendrites[1–3]. These dendrites can connect from the anode through the separator to the cathode (**Fig. 1**), thereby providing a low resistance path for electron transport (electrons will no longer flow through the external circuit and perform work), leading to high self-discharge currents that can ignite the flammable electrolyte, resulting in fires/explosions. This occurred when $Li_m$ was used as an anode in Moli Energy cells in the early 1990s and also in cells that were researched by Exxon around the same time[4]. Consequently, C replaced $Li_m$ in commercial cells while researchers continue to look for ways to solve the dendrite problem.

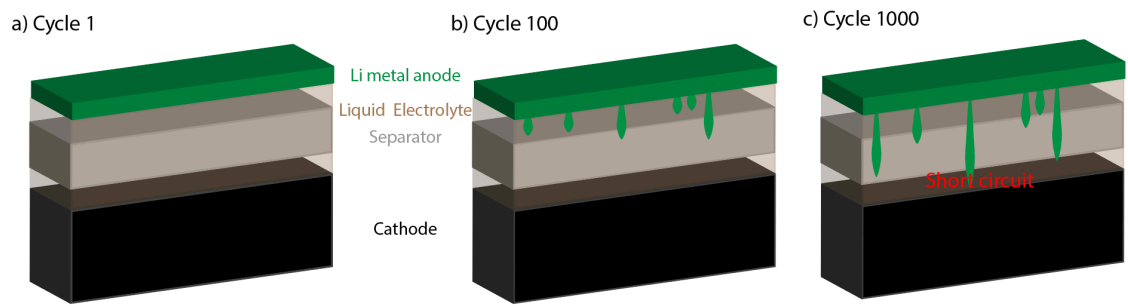

**Figure 1. Schematic of Li dendrite formation due to repeated charge and discharge cycles that could lead to a short circuit and ignite the flammable electrolyte.** Due to repeated deposition/stripping, Li metal begins to form dendrites that migrate through the separator. If the dendrites reach the cathode while still connected to the anode, a short circuit occurs that leads to rapid discharge and potential ignition of the flammable liquid electrolyte. In theory, a solid electrolyte prevents dendrites from reaching the cathode by physically blocking them. In practice, some solid electrolytes allow dendrite growth through grain boundaries. Several other methods, including electrolyte additives and nanostructuring, also show promise for eliminating dendrite growth in the presence of liquid electrolytes.

Solid state electrolytes (SSEs) are one of many approaches to solve the $Li_m$ dendrite problem[5]. In this approach, a SSE replaces the liquid electrolyte and acts as a physical barrier to dendrite penetration. Note, however, that some studies find dendrite penetration along grain boundaries in the SSE. The SSE is also generally less flammable than the liquid electrolyte in the event of a dendrite short and therefore safer[6]. While promising in this regard, SSEs are generally slow at transporting Li ions because ionic diffusion in a solid tends to be orders of magnitude slower than ionic diffusion in a liquid. Therefore, batteries that cycle with adequate rate capability are hard to build. This paradigm has been recently amended with the discovery of SSEs (from the thio-LISICON family) with conductivities higher than the standard LIB liquid electrolyte[7]. These discoveries have piqued the interest of many industrial companies and academic researchers. There have been several recent, comprehensive reviews that focus on solid state electrolytes[8–16], the practical challenges they face including processing and interfacial degradation[17], and particular classes of SSEs such as inorganic[9,18–22] and organic[23,24]. Additional reviews have focused on other strategies to enable the $Li_m$ anode[25,26] and how SSEs could enable beyond Li-ion chemistries such as Li-S and Li-$O_2$[27,28].

In this review, our focus is not on the structure-function relationships of different SSE families, strategies for improving their conductivity, or suggesting descriptors that may be used to find higher conductivity SSEs. Instead, we focus on recent reports of functioning solid state batteries (SSBs) and how they compare to current commercial LIBs. We also review commercial approaches to SSBs. Finally, we end with our perspective and timeline for future battery improvements.

**The LIB cost curve**

All future technologies that deviate from the oxide cathode/liquid electrolyte/carbon anode paradigm need to be evaluated with the following LIB cost curve in mind[29,30]:

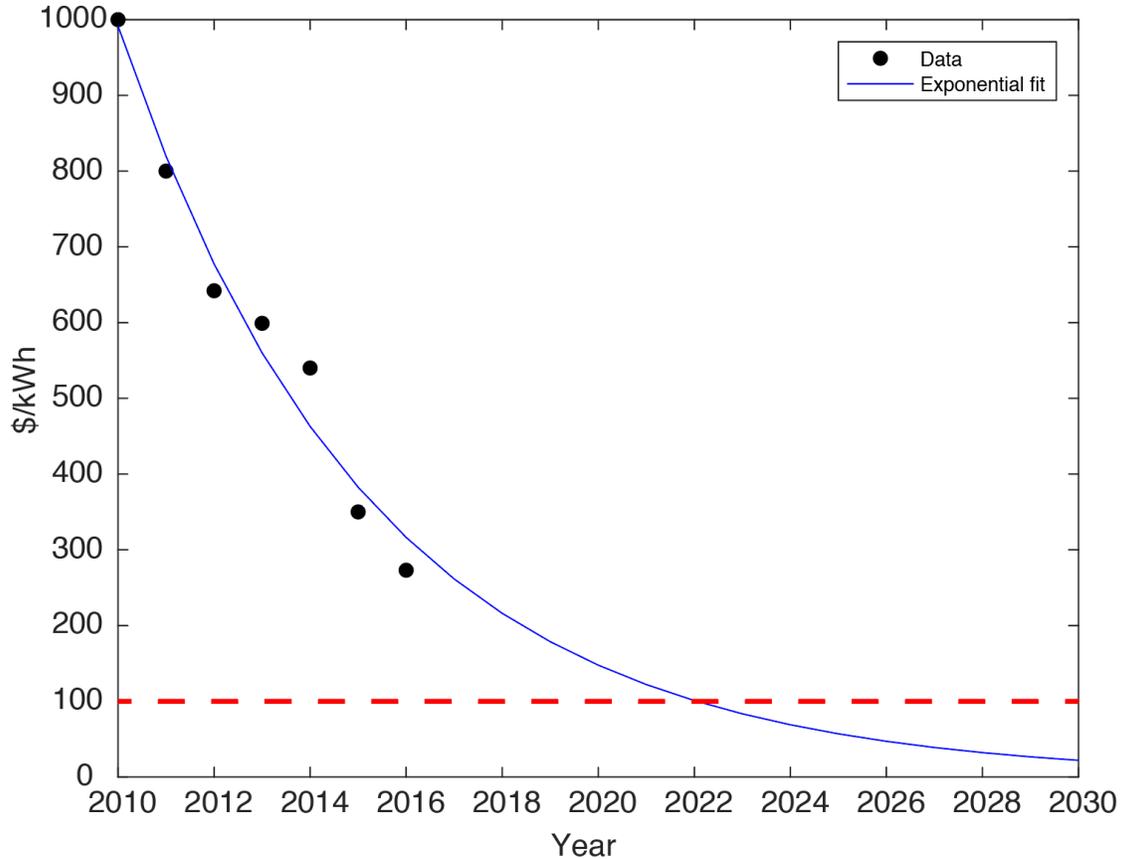

**Figure 2. Manufacturing cost of LIBs since 2010.** Data is shown in black and the exponential fit is shown in blue. Cost reduction has been driven by economies of scale, manufacturing learning rates, and improvements in cell chemistry and engineering. The so-called cost parity line of $100/kWh is drawn in red. It is generally accepted that the current chemistry can and will achieve cost parity.

The LIB cost curve (**Fig. 2**) is quite remarkable. The cost reduction is driven by many different factors, including cell manufacturing improvements, learning rates for pack integration, and capturing economies of scale (Gigafactories). The important thing to note is that cost parity with gasoline cars will likely be achieved within 5 years using currently technology. While we cannot discuss the commercial factors driving cost decreases, we will discuss how close current cell energy densities are to fundamental limits of the chosen chemistry. We will then compare these limits with those of a SSB.

**State of the art liquid electrolyte LIBs**

State of the art LIBs use oxide cathodes (particularly NMC and NCA)[31], liquid electrolytes with additives to improve coulombic efficiency[32–34], and carbon anodes with 3-5% silicon added to improve energy density[28]. A typical construction of a battery stack is shown schematically in **Fig. 3**.

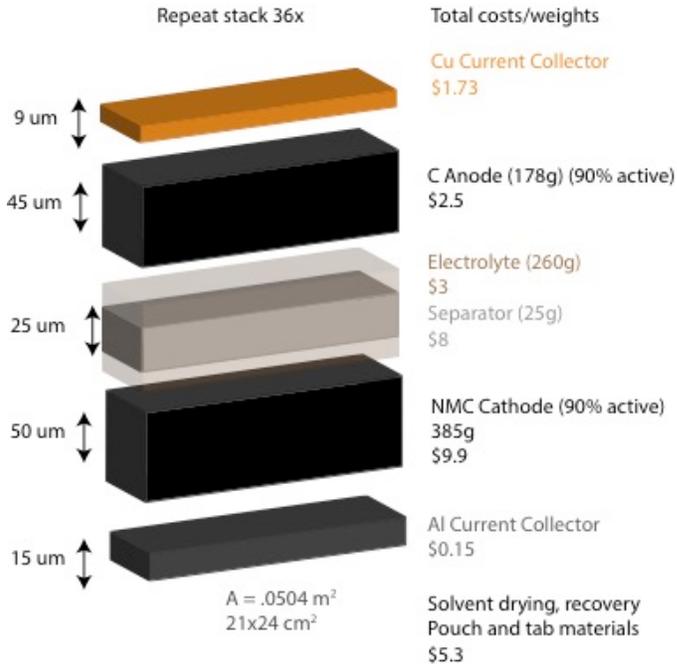

**Figure 3. A 52 Ah pouch cell stack with weights and costs of all of the components.** The total cost was estimated from the cost curve while the component percentage costs were estimated from previous literature[29,35,36]. In particular, we draw attention to the electrode active material percentages (90%), electrode thicknesses (~50 microns), and separator thickness (25 microns). A conceptually simple way to increase energy density is to increase electrode thicknesses thereby decreasing the weight contribution from the inactive material.

We first calculate the current energy density of commercially available LIBs. Assuming Tesla is using state of the art Panasonic batteries in their vehicles, the 100 kWh battery pack in the Model S P100D uses 8,256 18650 form factor cells[37], which is a total cell volume of 136.5 L leading to a volumetric energy density of 732 Wh/L. The weight of a single 18650 cell was not given in the teardown[37], but if we assume each one weighs about 45g then the gravimetric energy density is ~ 270 Wh/kg. The performance limit of this chemistry with the industry standard cathode thickness was recently estimated[2] to be 1200 Wh/L (400 Wh/kg). Note that if zero inactive components are used, the fundamental chemistry limit is ~ 470 Wh/kg. If the same cathode is used but the electrode thickness is increased by 60% and $Li_m$ replaces carbon as the anode, the performance is estimated[2] at 1300 Wh/L (475 Wh/kg). Note that increasing the electrode thickness increases energy density and decreases cost by decreasing the relative weight and volume contribution of the inactive materials[35]. The relative energy density increases due to $Li_m$ (+8% Wh/L,+19% Wh/kg) may seem surprising low as it is far from the naïve 10x increase expected from gravimetric specific capacity comparisons. To see why these numbers are lower than expected, consider the total cell gravimetric capacity definition:

$$Q \ (mAh/g) = \frac{m_c Q_c + m_a Q_a}{m_c + m_a}$$

where $m_c$ and $Q_c$ are the mass and gravimetric capacity (mAh/g) of the cathode, respectively, and $m_a$ and $Q_a$ are the equivalent values for the anode. This expression can be simplified using the capacity balance condition:

$$m_a = \frac{m_c Q_c}{Q_a}$$

to rewrite the specific cell capacity (excluding inactive components) as

$$Q \ (mAh/g) = \frac{1}{\frac{1}{Q_c} + \frac{1}{Q_a}}$$

Note that this is the same expression for the total resistance of two resistors in parallel. The specific cell energy is calculated by multiplying the above by the average voltage (U) and including the inactive mass ($m_{inact}$ (g/mAh)):

$$E \ (Wh/kg) = \frac{U}{\frac{1}{Q_c} + \frac{1}{Q_a} + m_{inact.}}$$

The point of this expression is to demonstrate the nonlinear relationship between anode gravimetric capacity and cell gravimetric energy density. Additionally, $Li_m$ has a relatively low bulk density (g/cm$^3$) relative to C that negatively impacts volumetric energy density. The same expression can be used to calculated E(Wh/L) by changing Q from mAh/g to mAh/mL by multiplying the values by the material density (g/cm$^3$).

Coulombic efficiency (discharge capacity/charge capacity x 100) is also incredibly important. Current LIBs exhibit average coulombic efficiencies (CEs) of > 99.99%, which is required for 74% capacity retention after 10 years of use (assuming 300 full charge/discharge cycles per year). If instead the average CE is 99.98%, which seems very similar, the capacity retention is only 55% after 10 years of use. Very small differences in CE thus result in very large differences in capacity retention after thousands of cycles. These metrics (Wh/L, Wh/kg, and CE) are now discussed for theoretical and actual SSBs.

**Theoretical solid state batteries**

The ideal solid state battery replaces the liquid electrolyte and separator with a solid electrolyte that is impenetrable to Li metal dendrites, thereby enabling Li metal as the anode. According to Ref. [2], SSBs using a Li metal anode can achieve ~ 480 Wh/kg using a low density SSE (such as a polymer or a sulfide) at a reasonable active material fraction (20% by Vol. inactive components). SSBs thus represent ~ 20% upside in energy density relative to current LIBs, but little upside relative to $Li_m$ liquid LIBs. Additionally, if a dense SSE is used, such as the garnet $Li_7La_3Zr_2O_{12}$, a theoretical energy density of 375 Wh/kg is possible, representing no upside to conventional liquid LIBs (**Table 1**). The main reason that the gains are less than expected from the $Li_m$ anode was discussed previously. Additional drawbacks for SSEs are the greater densities of SSEs relative to liquid electrolytes, the assumption of a 20 micron SSE thickness, and the lower active material fraction that is currently necessary to achieve reasonable cell conductivity in functioning SSBs[38].

**Table 1** compares the numbers for current and future liquid electrolyte and sold electrolyte LIBs.

| Technology | Gravimetric energy density (Wh/kg) | Volumetric energy density (Wh/L) | Cost ($/kWh) | Coulombic efficiency | Manufacturability |
|---|---|---|---|---|---|
| Current liquid electrolyte LIBs | 260 | 732 | 150 | 99.99% | Gigafactories operating |
| Future liquid electrolyte LIBs (no $Li_m$) | 400 | 1200 | 100 | 99.998% | Multiple gigafactories planned |
| Future liquid electrolyte LIBs with $Li_m$ | 475 | 1300 | ? | ? | ? |
| Current SSBs (low density SSE) | 155 | ? | ? | ~70% | ? |
| Future SSBs (low density SSE) | 480 | ? | ? | ? | ? |
| Future SSBs (high density SSE) | 375 | ? | ? | ? | ? |

**Table 1.** Comparing the current and future liquid and solid state LIB technology in terms of gravimetric and volumetric energy density, cost, coulombic efficiency, and manufacturability. Many values for liquid electrolyte $Li_m$ and SSBs with $Li_m$ are unknown as data is either not reported or unavailable.

Given the results in **Table 1** and calculations in Ref.[2], it's clear that low density SSEs are required to compete with current LIB technology. Given the current maximum conductivities of each family[9], we identify sulfides as the leading SSE technology, followed by polymers and then oxides. We discuss the performance metrics of SSBs using each class of SSEs reported in the literature, in addition to the promises and challenges of each.

**The Sulfides**

The sulfides, and in particular the thio-LISICON SSEs, were discovered in the early 2000s and are derived from LISICON (Lithium Super Ionic CONductor) compounds by replacing the oxygen with sulfur[39,40]. Realizations of the $Li_2S$-$GeS_2$-$P_2S_5$ family include LiGePS (LGPS), $Li_2S$-$P_2S_5$ (LPS), $Li_{9.54}Si_{1.74}P_{1.44}S_{11.7}Cl_{0.3}$ and others[7,41–46]. The lower electronegativity of sulfur relative to oxygen allows the movement of $Li^+$ ions more freely, resulting in higher conductivities relative to the LISICON family[7,41,43] (up to 10 mS/cm). These conductivities are sufficient to exceed the rate capabilities of current liquid electrolytes[42,47]. Additionally, several realizations of this family, including $Li_7P_3S_{11}$ (LPS), have sufficiently low densities (2 g/cm$^3$) such that the theoretical energy density with $Li_m$ represents significant upside to current LIBs[48]. While the conductivity and density of these materials are sufficient, significant stability and interfacial challenges exist[21,38]. In particular, sulfides are generally unstable both with respect to air (manufacturing challenge) and with respect to the electrode interface (both $Li_m$ and currently used oxide cathodes). A schematic of how SSBs using sulfides have been built in practice is given in **Fig. 4**.

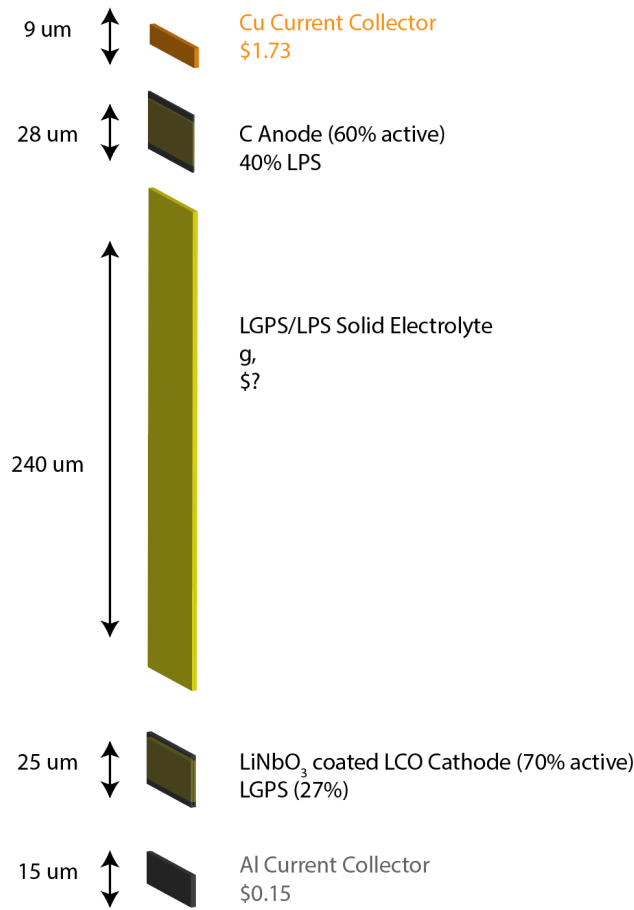

**Figure 4. Schematic of how Sulfide-based SSBs have actually been built.** Note the lower active material fractions (~60%), the thick SSE (240 microns), and the coating of the cathode material. These cell parameters would never compete with the state of the art LIBs shown in **Fig. 3**.

**Challenges and potential solutions**

There are two separate interfacial challenges. The first is generally true for all SSEs and is maintaining contact at a solid-solid interface. While the sulfides are more deformable than the oxides[49] (they can be densified by cold-pressing as opposed to high temperature sintering[50]), they still exhibit large interfacial impedances that prevent optimal cell operation. Indeed, practical cells require mixing the SSE with the cathode and anode powders to form composite materials in addition to using a thick SSE (see **Fig. 4**). Methods of addressing the solid-solid contact issue include using high pressure cells such that contact is maintained during the cycling-induced volume changes, preparing solution-processable SSEs that can then infiltrate the active material matrices thereby improving the number of conductive pathways[51], and adding a gel or liquid to form a hybrid mixture such that the liquid/gel accommodates the dynamic gap between the solids[52]. These approaches have yet to be used to build a SSB with a comparable CE to state of the art liquid LIBs.

The second interfacial challenge for the sulfides in particular is their reactivity with both the high voltage cathode and the $Li_m$ anode. For example, the thermodynamic stability window of LGPS is 1.7 to 2.14 V vs. $Li/Li^+$, indicating it will be unstable at both the Li metal anode, which is at 0 V relative to $Li/Li^+$, and also unstable at the 4 V or greater metal oxide cathode[21]. Bulk modifications to the SSE to improve its stability including doping the LGPS material with materials that interact strongly with $S^{2-}$, such as Ba, which was recently used to improve the high voltage stability[53]. However, this stability improvement was inferred from cyclic voltammetry sweeps, which can overestimate stability, and not from a functioning SSB. $MoS_2$ doping has also been used to improve the conductivity and stability of $Li_2S$-$P_2S_5$ glass-ceramic electrolytes[54]. In this case, the CE of a Li-S battery was improved due to the doping, although it still remained far from practical requirements. Interfacial modifications to the SSE have had the most success in terms of building functioning batteries. In these cases, the cathode material is coated with a thin oxide layer to reduce impedance and interfacial reactivity, while $Li_m$ is not used as the anode. Successful coatings include $ZrO_2$, $LiNbO_3$, LZO, and others[7,46,55–57].

An additional practical challenge is that the sulfides are generally not air stable. They react with moisture in the air to form $H_2S$ gas, which is toxic. However, the air stability can be modified through compositional changes. Recent work focused on minimizing $H_2S$ gas generation by changing the composition of sulfide glasses. $75Li_2S$-$25P_2S_5$ generated minimal gas[58]. Other approaches include the bulk addition of metal oxides[47].

**Functioning SSBs reported in the literature**

In **Table 2**, we review some of the functioning SSBs that have been reported in the literature. Note that many SSBs use coatings on the cathode[7,46,55] to mitigate the interface stability problems mentioned previously. Additionally, very few SSBs use $Li_m$ as the anode due to the aforementioned stability problems.

| Chemistry | Energy | Degradation | Year/Reference |

|  | density | rate |  |
|---|---|---|---|
| $Li_4Ti_5O_{12}$/LPS-LGPS/$LiCoO_2$ | 44 Wh/kg | Not reported | 2015[59] |
| Graphite/LGPS-LPS/$LiNbO_3$-coated $LiCoO_2$ | 180 Wh/kg, 435 Wh/L (thick electrode – 600 micon) | Not reported | 2018[60] |
| Graphite/$Li_6PS_5Cl$/NMC622 | 184 Wh/kg, 432 Wh/L | Cycled 20 times | 2018[61] |
| $Li_m$/LiI-$Li_3PS_4$/LZO-coated NCA | 260 Wh/kg (based on cathode composite, not including anode) | 99.6 within last 50 cycles | 2016[46] |
| Graphite/$Li_3PS_4$/$LiNbO_3$-coated NMC | 155 Wh/kg (excluding current collector and package) | Only reported for half cells | 2017[62] |
| Graphite/$Li_2S$-$P_2S_5$ (80:20 mol%)/$Li_2O$-$ZrO_2$ (LZO) coated NCA |  | 80% @ 100 cycles | 2014[57] |

**Table 2. Properties of SSBs built with sulfide SSEs.** Note that the reporting of gravimetric energy density is not always precise due to many studies omitting the inactive material weight. It is also common to quote capacities in mAh/$g_{cathode}$, which does not facilitate comparison to liquid LIBs and is arguably a misrepresentation of the actual performance.

As seen in Table 2, SSBs built with sulfide-based SSEs are improving but still lack the set of performance characteristics achieved with liquid electrolyte LIBs. That being said, we still believe sulfides represent the greatest potential opportunity due to their high conductivity and low gravimetric density.

**The Polymers**

Polymer SSEs are the only class of SSEs to be used commercially, including in pacemakers from Medtronic and the "BlueCars" from Bollore. The principle advantages are manufacturability and good interfacial contact. The principle disadvantage is the poor conductivity at room temperature, and the poor thermodynamic stability of some polymers. Consequently, commercial SSBs with polymer SSEs operate at elevated temperatures (37 C of the human body for pacemakers, 70-80 C by external heating in the Bollore cells) to achieve a reasonable conductivity.

**Approaches to enable higher conductivities**

General strategies for improving the conductivity of polymer electrolytes include synthesizing new polymer-salt complexes, utilizing soluble and insoluble additives[63], adding nanostructures[64,65], and forming hybrid electrolytes such as gels[12,23,24,66]. Recently, amorphous poly(ethylene ethere carbonate) was synthesized by a ring opening polymerization of ethylene carbonate[67]. The conductivity was $1.6 \times 10^{-5}$, which was sufficient to construct a room temperature SSB. This battery exhibited decent capacity retention (90% after 100 cycles) but also used a 200 micron thick $Li_m$ anode, which is certainly much more Li than required by simple capacity balancing[68] (probably 200% to several 1000%) and therefore would not achieve the energy densities discussed previously. A 3D cross-linked membrane approach was also recently used to build a $Li_m$/$LiFePO_4$ solid state battery[69]. The polymer SSE in this case was around 100 microns thick (negating some of the gravimetric energy density gains of $Li_m$) and the thickness of the $Li_m$ used was not reported.

**The Oxides**

One of the earliest lithium ion conductors with a crystalline structure was $Li_{14}ZnGe_4O_{16}$, which is a member of the LISICON family. This particular compound reaches a conductivity of $10^{-7}$ at room temperature, which is not sufficient for a SSB. However, when appropriately doped, oxides can reach conductivities as high as $10^{-3}$ S/cm (such as the garnet $Li_7La_3Zr_2O_{12}$ (LLZO)). The primary advantages of the oxides are their stability in air and better stability with respect to $Li_m$ and the high voltage cathodes. The primary drawbacks are their density, processability, manufacturability, and conductivity. LLZO is stable against Li metal and against a high voltage cathode. However, oxides generally require high sintering temperatures to remove grain boundaries to achieve the reported conductivity values. They also tend to be brittle, which makes it harder (relative to the sulfides) to maintain solid-solid interfacial contact and also to process. Strategies for improving the solid-solid interfacial issues include forming an intermediate Li-metal alloy that changes the wettability of the garnet surface[70], applying external pressure to the cell during cycling, using graded interfaces that were prepared via spark plasma sintering[71], mixing LLZO with a polymer to improve its flexibility[72], decreasing the interfacial impedance using coatings[73], and changing the garnet microstructure (grain boundaries, etc.). Finally, oxides tend to have high relative densities (g/cm$^3$) that may negate the benefit of moving to $Li_m$ in terms of energy density.

**Commercial approaches to enable Li metal**

We briefly review commercial approaches to enable $Li_m$ in the following sections.

**Inorganic approaches**

SolidPower, a spinoff of University of Colorado at Boulder, says it has made lab scale cells that reach 400-500 Wh/kg and up to 500 charge-discharge cycles[74,75]. According to the their website, the chemistry is $Li_m$/a high ionic conductivity solid separator (inorganic)/high capacity cathode. The stack is 100% inorganic. They have partnered with BMW and A123 Systems.

Toyota/AIST are pursuing both sulfides and oxides. Patents filed recently show they are interested in the single crystal garnet $Li_5La_3Ta_2O_{12}$ prepared in a novel way.

**Polymer approaches**

Medtronic uses the LiPON SSE with $Li_m$ in their pacemaker batteries. However, these batteries are extremely low capacity (.001 to .01 Ah) with extremely low discharge rates and the batteries are never recharged.

Cymbet offers 5-50 uAh devices at 3.8 V and is focused on standby power, portable devices, and the internet of things.

Ionic materials is using a SSE in an alkaline battery (Zn and $MnO_2$).

Thinergy is using the LiPON SSE and has cells with 2.2 mAh capacity.

Bollore, a European car company, is using a lithium-metal-polymer cell in their "blue cars"[76,75]. The car battery packs have a capacity of 30 kWh. However, this cell must be kept at 70-80 C and has an energy density of only 100 Wh/kg. They are using 3x the capacity balanced amount of Li, presumably due to the inefficiency of the stripping/deposition process.

SEEO reported in 2015 about their "drylyte" technology, which is a block copolymer SSE[77]. The capacity retention was 95% after 1000 cycles (100% depth of discharge, but no rate indicated). They claim 220 Wh/kg and demonstrated a 10 kWh pack. However, like other polymer-based systems, the batteries are held at high temperature that is thermally managed at the module level. They partnered with Bosch in 2015, since then there has been little news.

**Hybrid approaches**

Solid Energy, a spinout from MIT, claims to achieve 450 Wh/kg (1200 Wh/l) using a very thin Li metal anode, NMC cathode, and a solid protective coating consisting of polymer and inorganic materials that is applied directly to the surface-treated Li metal anode to suppress the growth of mossy lithium[78]. According to their website, they are currently producing 3 Ah cells. They have partnered with A123 Systems.

**Conclusions**

SSEs represent an opportunity to improve the gravimetric and volumetric energy density of LIBs by enabling the use of $Li_m$. However, many challenges remain and potential performance gains have, in many cases, been overpromised and oversold. Researchers should report active material weights, in particular the amount of $Li_m$ used, and inactive material weights, in particular the density and thickness of the SSE, for the correct comparison to current LIBs. In our view, the most promising SSE is a low density sulfide, such as LPS, or a low density polymer should one be found with sufficient room temperature conductivity. Of course, there are applications in which SSEs confer significant advantages, such as rapid charge/discharge (rates greater than 10 C), extreme temperature performance (greater than 80 °C or less than 10 °C), and improved safety. These applications are important but of much less scale than automotive, consumer electronic, and stationary storage applications. Ultimately, the technical challenges with SSEs coupled with the rapidly decreasing cost of liquid electrolyte LIBs leads us to conclude that the current LIB paradigm will likely continue for the foreseeable future.

**Appendix**

**Model 3 battery cost**. If we assume the cost to Tesla is $190/kWh then the 55 kWh model battery costs approximately $10,450. If we assume a 20% gross margin and a $35,000 base model price then this implies the total cost to Tesla is $28,000. This means the battery is approximately 37% of the total cost.